\documentstyle[aasms4]{article} 

\newcommand{\w}{\tilde{w}} 
\newcommand{\p}{\bar{p}}
\newcommand{\W}{\tilde{W}} 
\newcommand{\tdelta}{\tilde{\delta}}
\newcommand{\lfac}{{ 2 \ell +1 \over 4\pi}}
\newcommand{\lpfac}{{ 2 \ell' +1 \over 4\pi}}
\newcommand{\pllint}{{2\ell + 1 \over 4 \pi}{2\ell' + 1 \over 4 \pi}
\int d\Omega_{K} d\Omega_{K'}{\cal 
P}_{\ell}(\mu_{K}){\cal P}_{\ell'}(\mu_{K'})
}

\begin{document}
\title{Sensitivity of Redshift Distortion Measurements to  Cosmological
Parameters}

\author{Andrew A.\ de Laix and Glenn Starkman}

\affil{Physics Department, Case Western Reserve University, Cleveland,
OH 44106-7079}

\authoremail{aad4@po.cwru.edu}

\begin{abstract} 
The multipole moments of the power spectrum of large scale structure,
observed in redshift space, are calculated for a finite sample volume
including the effects of both the linear velocity field and geometry.
A variance calculation is also performed including the effects of shot
noise.  The sensitivity with which a survey with the depth and
geometry of the Sloan Digital Sky Survey (SDSS) can measure
cosmological parameters $\Omega_0$ and $b_0$ (the bias) or $\lambda_0$
(the cosmological constant) and $b_0$ is derived through fitting power
spectrum moments to the large scale structure in the linear regime in
a way which is independent of the evolution of the galaxy number
density.  A fiducial model is assumed and the region of parameter
space which can then be excluded to a given confidence limit is
determined.  In the absence of geometric and evolutionary effects, the
ratios of multipole moments (in particular the zeroth and second), are
degenerate for models of constant $\beta \approx \Omega^{0.6}/b_0$.
However, this degeneracy is broken by the Hubble expansion, so that in
principle $\Omega_0$ and $b_0$ may be measured separately by a deep
enough galaxy redshift survey (\cite{nakamura}).  We find that for
surveys of the approximate depth of the SDSS no restrictions can be
placed on $\Omega_0$ at the 99\% confidence limit when a fiducial
open, $\Omega_0 = 0.3 $ model is assumed and bias is unconstrained.
At the 95\% limit, $\Omega_{0} < .85$ is ruled out.  Furthermore, for
this fiducial model, both flat (cosmological constant) and open models
are expected to reasonably fit the data.  For flat, cosmological
constant models with a fiducial $\Omega_{0} = 0.3$, we find that
models with $\Omega_{0} > 0.48$ are ruled out at the 95\% confidence
limit regardless of the choice of the bias parameter, and open models
cannot fit the data even at the 99\% confidence limit.  We also find
significant deviations in $\beta$ from the naive estimate for both
fiducial models.  Thus, we conclude for the SDSS that linear
evolution-free statistics alone can strongly distinguish between
$\Omega_0 = 1$ and low matter density models only in the case of the
fiducial cosmological constant model.  For the open model, $\Omega_{0}
= 1$ is only at best only nominally excluded unless $\Omega_0 < 0.3$.

\end{abstract} 


\section{Introduction}

With the expectation of new and much larger samples of galaxies with
measured redshifts, {\it e.g.}~the Sloan Digital Sky Survey (SDSS, see
{\it e.g.}~\cite{gunn}), it is opportune to reconsider what
information may be extracted from these data.  In particular, we would
like to determine how well fundamental cosmological parameters like
the current matter density $\Omega_0$ and cosmological constant
$\lambda_0$ can be inferred from redshift surveys.  Perhaps the
simplest approach would be to just measure the mean number density of
galaxies as a function of redshift.  Unfortunately, this number
density is evolving through processes other than cosmological
expansion, and these factors must be accurately deconvolved 
if we are to extract the desired information about cosmology.  
Alternatively, in the standard cosmological model, 
the growth of perturbations in galaxy number density 
is driven only by gravity, 
so we expect that measurements derived from these variations 
will be cleaner signals of cosmology.  
Locally, this perturbation growth induces peculiar-velocities in the galaxies 
which lead to distortions in redshift maps.  
At increasing redshift, the effects of these distortions evolve 
due to the changing growth rate of perturbations,
and due to the geometry and expansion of the universe.  
This evolution, in turn, depends on fundamental cosmological parameters
like the curvature, $\Omega_0$, and $\lambda_0$, 
so one may hope to infer values of these parameters 
by observing the change in the distortions with redshift. 
\cite{nakamura} have recently calculated the effects  
on the expectation value of the various multipoles of the correlation function,
to first order in the redshift. 
However, they made no estimate of the expected errors, 
necessary for determining the sensitivity of any real survey.  

In this paper we shall derive the mean, cosmic variance and shot noise
associated with the statistics of multipole moments of the linear
power spectrum calculating the redshift-dependent effects to all
orders.  Note that moments beyond the zeroth are induced by redshift
distortions and would not be present in the mean for a real-space
survey, {\it i.e}.\ a survey which plots galaxies at their conformal
distances.  Then, we estimate the sensitivity of these statistics in
the linear regime for a survey with depth and geometry comparable to
the SDSS by determining those models which can be excluded to given
confidence level when a particular fiducial model is assumed.  In this
way, we hope to put into perspective the ability of upcoming redshift
surveys to fix cosmological parameters in the absence of external
constraints.

\section{Redshift Distortions in The Power Spectrum} 
\label{distortion}
The theory of the linear distortions of redshift surveys with regards
to the underlying real--space distribution was first investigated by
\cite{kaiser} for shallow surveys where redshift evolution and
geometrical effects were negligible.  This work has been extended to
include both evolution and geometry for larger redshifts by
\cite{matsubara} and \cite{ballinger}; for convenience and clarity, 
we shall repeat their work following the formalism of Matsubara and
Suto.  We begin with the standard assumption of a
Friedman-Robertson-Walker universe with a metric given by
%
\begin{equation}
\label{metric}
ds^{2} = -dt^2 + a(t)^2 \{ d \chi^2 + S(\chi)^2(d\theta^2 +\sin^2\theta
~d\phi^2)\},
\end{equation}
where $S(\chi)$ is determined by the geometry of the universe through
the spatial curvature $K$, and 
%
\begin{equation}
\label{S}
S(\chi) = \left\{
	\begin{array}{lr}
		\sin(\sqrt{K}\chi)/\sqrt{K} & (K > 0), \\
		\chi & (K = 0), \\
		\sinh(\sqrt{-K}\chi)/\sqrt{-K} & (K < 0).	
	\end{array}
\right.	 
\end{equation}
The curvature is given in terms of the present day Hubble constant
$H_{0}$, the matter energy density $\Omega_{0}$ and the energy density
in cosmological constant $\lambda_{0}$:
%
\begin{equation}
\label{curve}
K = H_{0}^2(\Omega_{0} + \lambda_{0} -1),
\end{equation}
where we assume that the present scale factor $a_0$ is unity.  The
proper radial distance from an observer to a source,
$\chi$, can be determined from the integral 
%
\begin{equation}
\label{chi}
\chi = \int^{t_0}_{t}{dt\over a(t)} = \int^{z}_{0} {dz \over H(z)} ,
\end{equation}
where in the final integral, we introduce the redshift--dependent
Hubble parameter	
%
\begin{equation}
\label{hubble}
H(z) = H_0 \sqrt{\Omega_0 (1+z)^3 + (1 - \Omega_0 - \lambda_0) (1+z)^2
+ \lambda_0}.
\end{equation}

To see how redshift-space maps are distorted with regards to their
real--space counterparts, we need to consider both the geometry of the
universe and the peculiar--velocities of the objects being observed.
Let us first consider geometry.  We would like to examine small
displacements about a given origin located at a redshift $z$ with
respect to a terrestrial observer ($z=0$).  This small displacement is
represented by the vector $\vec{x}$ in comoving, real--space
coordinates and has components $x_\|$ parallel to the line--of--sight
and $x_\bot$ perpendicular to the line--of--sight.  To first order in
the Taylor expansion we may write
%
\begin{eqnarray}
\label{xofz}
x_\| &=& {d\chi(z) \over dz} \delta z = {c_\| \over H_0} \delta z
\nonumber \\
\vec{x}_\bot &=& S(\chi(z)) \delta \vec{\theta} = {c_\bot z \over H_0} \delta
\vec{\theta},
\end{eqnarray}
where $c_\| = H_0/H(z)$ and $c_\bot = H_0 S(\chi(z))/z$. We have also
assumed a distant observer by linearizing in $\delta \theta$.  Since
$c_\|$ and $c_\bot$ have different functional forms, a spherical
object will be distorted into an ellipse when mapped into redshift
space, and it is this effect that \cite{alcock} suggested could be
used to measure cosmological parameters.

For galaxy surveys we also need to consider how the peculiar
velocities of objects affect redshift distributions.  A galaxy 
that would be observed at redshift $z_{true}$ in the absence of any
peculiar motion, and which has a non-relativistic peculiar velocity
$\vec{v}$ relative to the background, is seen by an observer
with non-relativistic peculiar velocity $\vec{v}^0$ with an apparent
redshift
%
\begin{equation}
\label{zobs}
1 + z_{app} = (1 + z_{true})(1 + v_\| - v_\|^0).
\end{equation}
Here $v_\|$ is the peculiar velocity of the object projected along the
line--of--sight.  Using eq. (\ref{zobs}) along with the
eq. (\ref{xofz}), we can calculate the displacement in redshift-space
$\vec{s}$ for a given $\vec{x}$ including both the geometric and
velocity effects. If we drop the term proportional to $H x_\| (1 +
v_\| - v_{\|}^O)$, which is small in comparison to the
others, we can easily show that
%
\begin{eqnarray}
\label{sdef}
s_1 &=& {x_1 \over c_\bot(z)},~s_2 = {x_2 \over c_\bot(z)}, \nonumber
\\ s_3 &=& {z_{app} - z \over H_0} \simeq {1 \over c_\|(z)}\left[ x_3 +
{1 + z \over H(z)}(v_\| - v_{\|}^O) \right],
\end{eqnarray}
where we recall that $z$ is the redshift of the origin of reference
and $z_{true} - z$ is equivalent to $\delta z$ in eq.\ (\ref{xofz}). Note
that the 3 direction points along the line--of--sight.  The above
expressions now give us almost everything we need to know to relate
the real--space density--contrast to the redshift-space contrast. 
The conversion is made by considering the Jacobian transformation from
$\vec{x}$ to $\vec{s}$:
\begin{equation}
\label{jacobian}
\left|{\partial \vec{x} \over \partial \vec{s}} \right| 
\approx c_\bot^2 c_\|\left[
1 - {1+z \over H(z)} {\partial \over \partial x_3}
v_\|(\vec{x})\right], 
\end{equation}
to linear order in the velocity perturbation, where the local density
in real space $\delta^r(vec{x})$ is thus enhanced by this factor when
observed in redshift space. For the mean density, we take the ensemble
average, so the velocity term vanishes when averaging the Jacobian.
Putting everything together, we get
%
\begin{equation}
\label{deltas}
\delta^{s}(\vec{s}(\vec{x})) = \delta^{r}(\vec{x}) - {1+z \over H(z)}
{\partial \over \partial x_3} v_\|(\vec{x}). 
\end{equation}
This result is true for any fluid component, {\it e.g.}~galaxies, dark
matter, etc., as it depends only on the continuity of the fluid, but
in the special case of the {\it total mass} fluctuations $\delta^m$ we
can go further and replace the velocity term by assuming, in the
linear regime, that fluctuations grow only by gravitational
instability. Perturbation theory gives us a relationship between
$\delta^m$ and the peculiar velocity, namely
%
\begin{equation}
\label{deltav}
v_\|(\vec{x}) = -{H(z) \over 1+z}f(z) \partial_3 \triangle^{-1}
\delta^m(\vec{x}),
\end{equation}
where $\triangle^{-1}$ is the inverse of the Laplacian operator (see
{\it e.g.} Peebles 1980) and $f(z)$ is the logarithmic derivative of
the linear growth rate $D(z)$,
%
\begin{equation}
\label{f}
f(z) = {d \ln D(z) \over d \ln a} \simeq \Omega(z)^{0.6} + {\lambda(z)
\over 70} \left[ 1 + {\Omega(z) \over 2} \right ].
\end{equation}
In this equation, the redshift dependent cosmological parameters
$\Omega(z)$ and $\lambda(z)$ are given by 
%
\begin{equation}
\label{OmegaLambda}
\Omega(z) = \left[ {H_0 \over H(z)} \right]^2 (1+z)^3 \Omega_0,~
\lambda(z) = \left[ {H_0 \over H(z)} \right]^2 \lambda_0,
\end{equation}
(during the matter-dominated epoch)
while the linear growth factor has the well known solution (\cite{peebles80}) 
%
\begin{equation}
\label{growth}
D(z) = {5 \Omega_0 H_0^2 \over 2} H(z) \int_{z}^{\infty} {1+z' \over
H(z')^3} d z'.
\end{equation}

Galaxies do not necessarily trace mass, but, to first order we may
assume that the relationship between galaxy and mass fluctuations is
linear, given by a bias factor $\delta^r = b(z) \delta^m$.  The
redshift dependence of this bias factor can be quite complex,
depending on the dynamics of galaxy formation, but for $z \ll 1$ it is
reasonable to follow \cite{fry} and suppose that all the galaxies in
our survey were formed well prior to their observation with some
intrinsic bias.  Treating these galaxies as a separate matter
component in the perturbation equations, Fry has shown that
%
\begin{equation}
\label{bias}
b(z) = 1 + {D(0) \over D(z)}(b_0 - 1).
\end{equation}

Putting everything together, we may now write the linear overdensity
function as a function of $\vec{s}$ in the neighborhood of a given
origin at $z$:
%
%
\begin{equation}
\delta^s(\vec{s}(\vec{x})) = {\delta}^r(\vec{x}) + \beta(z) {\partial^2 
\over {\partial x_3}^2} \triangle^{-1} {\delta}^r(\vec{x}),
\end{equation}
where $\beta(z) = f(z)/b(z)$.  Both for expressing the density
fluctuations, usually assumed to be a Gaussian random field, and for
evaluating the inverse Laplacian, it is convenient to work in Fourier
rather than real space.  Thus we can write
%
\begin{equation}
\label{deltaLin}
\delta^s(\vec{s}(\vec{x})) = \int {d^3k \over (2\pi)^3} \left[ 1 +
\beta(z) {k_3^2 \over k^2} \right] {b(z) D(z) \over b(0) D(0) }
\tilde{\delta}^r(\vec{k}) e^{i\vec{k}\cdot\vec{x}(\vec{s})},
\end{equation}
where $\tilde{\delta}(\vec{k})$ is the Fourier
transform of the real--space density contrast evaluated at $z=0$.

The above result is more than adequate for perturbations on the
largest scales, but on smaller scales, we must account for the
non--linear evolution.  This causes the density fluctuations to
rapidly virialize into halos, and the effect on the redshift--space
density field is well approximated by adding a random velocity to all
the mass elements.  Empirical evidence suggests that the distribution
of these velocities is well described by a power--law model with
dispersion $\sigma_{v}$ (\cite{cole}).  Thus, the corrected density
fluctuation in redshift space then takes the form
\begin{equation}
\label{deltaF}
\delta^s(\vec{s}(\vec{x})) = \int {d^3k \over (2\pi)^3} \left[ 1 +
\beta(z) {k_3^2 \over k^2} \right] \left[1+{k_{3}^{2}\sigma_{v}^{2} 
\over 2}\right]^{-1}
{b(z) D(z) \over b(0) D(0) }
\tilde{\delta}^r(\vec{k}) e^{i\vec{k}\cdot\vec{x}(\vec{s})}.	
\end{equation}
Now let us consider the Fourier coefficients $\tdelta^{s}(\vec{K})$,
which we would derive from the redshift-space density distribution, 
defined as
\begin{equation}
\label{deltaS}
	\tdelta^{s}(\vec{K }) \equiv \int d^{3}s 
	~\delta^s(\vec{s}(\vec{x}))  
	e^{-i\vec{K}\cdot \vec{s}}.
\end{equation}
Substituting in the definition of $\delta^s(\vec{s}(\vec{x}))$ from 
eq.\ (\ref{deltaF}), we get
\begin{equation}
\label{deltaS1}
	\tdelta^{s}(\vec{K }) = \int d^{3}s 
	\int {d^3k \over (2\pi)^3} \left[ 1 +
        \beta(z) {k_3^2 \over k^2} \right] \left[1+{k_{3}^{2}\sigma_{v}^{2} 
        \over 2}\right]^{-1}
        {b(z) D(z) \over b(0) D(0) }
        \tilde{\delta}^r(\vec{k}) e^{i\vec{k}\cdot\vec{x}(\vec{s})}
	e^{-i\vec{K}\cdot \vec{s}}.
\end{equation}
Recall that $\vec{x}(\vec{s}) = c_{\bot} \vec{s}_{\bot} + 
c_{\|} \vec{s}_{\|}$, so by performing the integration over 
$d^{3}s$ we are left with 
 \begin{equation}
 \label{deltaS2}
 	\tdelta^{s}(\vec{K }) = \int {d^3k} \left[ 1 +
        \beta(z) {k_3^2 \over k^2} \right] \left[1+{k_{3}^{2}\sigma_{v}^{2} 
        \over 2}\right]^{-1}¥
        {b(z) D(z) \over b(0) D(0) }
        \tilde{\delta}^r(\vec{k}) \delta^{D}(\vec{K} - 
	c_{\bot} \vec{k} _{\bot} + c_{\|} \vec{k}_{\|}),
\end{equation} 
where $\delta^{D}(\vec{K} - c_{\bot}\vec{k}_{\bot} + c_{\|} 
\vec{k}_{\|})$ is a Dirac delta function. The final integration is now 
trivial to perform, and doing so leaves us with
\begin{eqnarray}
\label{deltaSF}
 	\tdelta^{s}(\vec{K }) &=& \left[ 1 + {\beta \mu^{2} c_{\|}^{-2} 
 	\over \left( c_{\|}^{-2} -c_{\bot}^{-2}\right)
 	\mu^{2}+c_{\bot}^{-2}} \right] 
 	\left(1 + {1 \over 2} c_{\|}^{-2} 
 	\sigma_{v}^{2}K^{2}\mu^{2}\right)^{-1} \\ \nonumber 
 	&& \times \tilde{\delta^{r}}\left( K \sqrt{(c_{\|}^{-2} 
	-c_{\bot}^{-2}) 	
 	\mu^{2}+c_{\bot}^{-2}}\right)\left(c_{\|}c_{\bot}^{2}
	\right),
\end{eqnarray}
where $\mu = K_{3}/K$.  

\section{Fluctuations in Galaxy Counts}
In the previous section, we derived an expression for the Fourier
components of the galaxy density field mapped into redshift space
valid in the neighborhood of some observation point.  Unfortunately,
the results fail as $\vec{s}$ grows large enough to break the
linearization constraints enforced to calculate $c_\|$ and $c_\bot$,
so we can only apply them to small volumes.  The question becomes, how
can we best use these results to analyze redshift surveys which cover
large angles of sky and are deep in redshift space?  The only
practical alternative is to break up such a survey into many
sub--volumes, and measure the distortions within each of them.  These
measurements can then be fit to theoretical calculations and used to
determine cosmological parameters.  In \S \ref{momentssec}, we shall
discuss the calculation of multipole moments of the power spectrum as
observed in a single sub--volume, while in \S \ref{covar} we will
calculate the cosmic variance of these moments, for a single volume.
Finally, in \S \ref{shotNoise}, we calculate the shot noise or finite
sampling variance for this sub--volume.  Together, these computations
will allow us to determine the sensitivity of redshift surveys which
contain many statistically independent sub--volumes.

\subsection{Moments of the Galaxy Distribution}
\label{momentssec}
The mean value for multipole moments of the power spectrum derived
from a finite sample window were first calculated in \cite{cole0};
here we repeat their calculation, to which we will add a calculation
of the variance of these statistics in the following subsections.
Since we are considering galaxy surveys as our source of data, we must
consider not continuous density fields, but discrete realizations of
those fields given by the point locations of the galaxies.  This is a
well understood problem, so we shall follow
\cite{peebles80}, and subdivide our survey into a large number of
cells such that the probability of finding two galaxies in a cell is
vanishingly small in comparison to that of finding one in a cell.  We
define $N_{i}$ to be the number of galaxies in cell $i$, where by our
supposition $N_{i}$ is either one or zero.  In the previous section,
we considered the Fourier components of the density fluctuations in a
neighborhood about a fixed redshift.  For a corresponding survey, we
ought to only consider galaxies inside a window such that the
linearization performed previously is a valid approximation, and thus
we will weight each of our cells by a window function $w(\vec{s}_{i})$
with volume $V_{w}$.  Also, real galaxy surveys usually do not
uniformly sample all the volume in the survey, {\it e.g.}~a flux
limited survey sees fewer and fewer objects as distance increases; the
function $\phi(\vec{s}_{i})$ will represent this selection
effect. Putting everything together, we see that a sensible way to
calculate the Fourier coefficients of the density fluctuations is
\begin{equation}
\label{deltaK}
	\tdelta_{\vec{K}} = {1 \over n V_{w}} \sum_{i} \left( N_{i}-\langle 
	N_{i}\rangle \right) {w(\vec{s}_{i}) \over \phi(\vec{s}_{i})} 
	e^{-i\vec{K}\cdot \vec{s}_{i}},
\end{equation}
where $n$ is the mean number density in the volume and angle brackets
refer to the ensemble average.  Dividing by $\phi$ will, as we shall
see, cancel its effects on the statistics we calculate rendering
selection-independent results.  The expectation--value for the number
of galaxies in cell $i$ is given by $\langle N_{i}\rangle = n
\phi(\vec{s}_{i}) d^{3}s_{i}$.  As we shall see momentarily, dividing
the $\tdelta_{\vec{K}}$'s by the selection function will remove 
the effects of the latter from the statistics we calculate.

In the previous section we showed that the Fourier coefficients of the
density field in redshift space are dependent on $\mu$, the cosine of
the angle formed by $\vec{K}$ and the observation axis.  Furthermore,
the angular dependence is a function of cosmology which appears
through $\beta$, $c_{\|}$ and $c_{\bot}$.  One might hope\cite{nakamura}
that by measuring the angular dependence of the $\delta_{\vec{K}} $'s 
one can
determine the underlying cosmology, and thus, to that end, we consider
the multipole moments of the square of the Fourier coefficients
\begin{equation}
\label{PL}
	\bar{p}_{\ell}(K) \equiv \lfac \int d\Omega_{K}{\cal
	P}_{\ell}(\mu_{K}) \left\langle
	\tdelta(\vec{K})\tdelta(\vec{K})^{*} \right\rangle,
\end{equation}
where ${\cal P}_{\ell}$ is a Legendre polynomial.  We choose the square 
of the Fourier coefficient because the ensemble average of the 
coefficient itself is zero and therefore we would see no signal in the 
mean.  Substituting in for $\tdelta_{\vec{K}}$ from eq.\ 
(\ref{deltaK}) and taking the ensemble average, we find
\begin{equation}
\label{PL1}
	\bar{p}_{\ell}(K) = \lfac \int d\Omega_{K}{\cal
	P}_{\ell}(\mu_{K}){1 \over n^{2}V_{w}^2} \sum_{i,j}\langle
	(N_{i}-\langle N_{i} \rangle ) (N_{j}-\langle N_{j} \rangle )
	\rangle {w(\vec{s}_{i}) w(\vec{s}_{j}) \over \phi(\vec{s}_{i})
	\phi(\vec{s}_{j})} e^{i\vec{K}\cdot(\vec{s}_{i}- \vec{s}_{j})}
\end{equation}
The remaining expectation--value can be written 
\begin{equation}
\label{NN}
	\langle (N_{i}-\langle N_{i} \rangle )
	(N_{j}-\langle N_{j} \rangle ) \rangle = 
	n^{2}\phi(\vec{s}_{i})\phi(\vec{s}_{j})d^{3}s_{i}d^{3}s_{j}\xi_{ij}
	+ n \phi(\vec{s}_{i})d^{3}s_{i}\delta_{{ij}},
\end{equation}
where $\delta_{ij}$ is the Kroniker delta, $\xi_{ij}$ is the
correlation function given by
\begin{equation}
\label{xi}
	\xi_{ij} \equiv \int {d^{3}k \over (2\pi)^{3}} e^{i\vec{k}\cdot 
	(\vec{s}_{i}-\vec{s}_{j})} P(\vec{k}),
\end{equation}
and $P(\vec{k}) = \tdelta^{s}(\vec{k})\tdelta^{s*}(\vec{k})/(2\pi)^{3}$ 
is the redshift--space power--spectrum derived from the perturbation 
spectrum given in eq.\ (\ref{deltaSF}).  It may be conveniently
expressed as an expansion in multipoles
\begin{equation}
\label{Pl}
P(\vec{k}) = \sum_{\ell} {\cal P}_\ell(\mu_k) P_\ell(k),
\end{equation}
where $\mu_k = k_3/k$ is the cosine of the angel formed between
$\vec{k}$ and the line of sight.  Putting everything together, we find
\begin{eqnarray}
 \label{PB} \bar{p}_{\ell}(K) &=& \lfac \int d\Omega_{K}{\cal
 P}_{\ell}(\mu_{K}) {1 \over n^{2}V_{w}^2} \sum_{i,j} \left[
 \vphantom{{w(\vec{s}_{i}) w(\vec{s}_{j}) \over \phi(\vec{s}_{i})
 \phi(\vec{s}_{j})}} n^{2}\phi(\vec{s}_{i})\phi
 (\vec{s}_{j})d^{3}s_{i}d^{3}s_{j} \right.  \\ \nonumber && \left
 .\times \int {d^{3}k \over (2\pi)^{3}} e^{i\vec{k}\cdot
 (\vec{s}_{i}-\vec{s}_{j})} P(\vec{k}) + n
 \phi(\vec{s}_{i})d^{3}s_{i}\delta_{{ij}}\right] {w(\vec{s}_{i})
 w(\vec{s}_{j}) \over \phi(\vec{s}_{i}) \phi(\vec{s}_{j})}
 e^{i\vec{K}\cdot(\vec{s}_{i}- \vec{s}_{j})}.
\end{eqnarray} 
In the limit of infinitesimal
volumes, the sums over $i$ and $j$ reduce to integrals, and we
see in the first term that these spatial integrals give us
Fourier transforms, reducing the above equation to
\begin{equation} 
\label{PB1} 
	\bar{p}_{\ell}({K})  =  \lfac \int d\Omega_{K}{\cal
	P}_{\ell} (\mu_{K}) \int {d^{3}k \over (2\pi)^{3}} P(\vec{k})
	\left|\w^{(1)}(\vec{K}+\vec{k})\right|^{2}  + {
	\delta_{\ell 0} \over n V_{w}^{2}} \int d^{3}s
	{\left|w(\vec{s})\right|^{2} \over \phi(\vec{s})},
\end{equation} 
where
\begin{equation}
\label{wp}
	\w^{(p)}(\vec{k}) \equiv \int {d^{3}s \over V_{w}} w^p(\vec{s}) 
e^{i \vec{k}\cdot \vec{s}}.
\end{equation}
To further reduce the integral in eq.\ (\ref{PB1}), we shall assume 
that our window function is spherically symmetric implying
$\w(\vec{k}) = \w^{*}(\vec{k}) = \w(|\vec{k}|)$.  We can then write out 
the multipole expansion 
\begin{equation}
	\left|\w^{(1)}(\vec{K}+\vec{k})\right|^{2}  = 
	\left[\w^{(1)}(\vec{K}+\vec{k})\right]^{2} 
\equiv \W^{(1,1)}(\vec{K}+\vec{k}) \equiv 
	\sum {\cal P}_{n}(\mu_{K,k})
	\tilde{W}^{(1,1)}_{n}(K,k),	
\end{equation}
where $\mu_{K,k}$ is the cosine of the angle formed between $\vec{K}$ 
and $\vec{k}$.  
More generally, we will define 
\begin{equation}
\label{Wpq}
\W^{(p,q)}(\vec{K}+\vec{k}) \equiv 
\w^{(p)}(\vec{K}+\vec{k}) \w^{(q)}(\vec{K}+\vec{k}) 
\end{equation}
and use the multipole expansions
\begin{equation}
\label{wnp}
\w^{(p)}(\vec{K}+\vec{k}) \equiv 
	\sum {\cal P}_{n}(\mu_{K,k})
	\tilde{w}^{(p)}_{n}(K,k).
\end{equation}
and 
\begin{equation}
\label{Wnpq}
\W^{(p,q)}(\vec{K}+\vec{k}) \equiv 
	\sum {\cal P}_{n}(\mu_{K,k})
	\tilde{W}^{(p,q)}_{n}(K,k).
\end{equation}
As we have already mentioned (eq.\ (\ref{Pl})), $P(\vec{k})$ can also 
be written as a multipole expansion in $\mu_{k}$.  To make use of 
these expansions, we will apply a well known property of Legendre 
polynomials, namely
\begin{equation}
\label{LL}
	\int_{}^{} d\Omega_{k}{\cal P}_{n}(\mu_{K,k}){\cal P}_{m}(\mu_{k}) =
	{4\pi \over 2n+1}{\cal P}(\mu_{K})\delta_{nm}.
\end{equation}
Using this theorem along with the expansions of the window function 
and power spectrum, one can see after some algebra that 
\begin{eqnarray}
\label{PBfinal}
	 \bar{p}_{\ell}({K}) & = &  {4 \pi \over 2 \ell 
	 +1} \int_{0}^{\infty} {k^{2}dk \over (2 \pi)^{3}} 
	 P_{\ell}(k) \tilde{W}^{(1,1)}_{\ell}(K,k)
	  + { \delta_{\ell 0}
 	 \over n V_{w}^{2}} \int d^{3}s {\left|w(\vec{s})\right|^{2}
 	 \over \phi(\vec{s})},
\end{eqnarray}
where the second term vanishes as the number of galaxies in the survey
volume grows large. Conversely, it is easily removed from measurements in 
surveys for which shot noise makes a significant contribution.

\subsection{Variance of the Moments}
\label{covar}
Now that we have derived the multipole moments of the power
spectrum for a finite volume, one can imagine taking many such
volumes in a redshift survey and calculating $ \bar{p}_{\ell}$'s as a
function of $z$.  Then by using the theoretical results, one could fit
various cosmological models to the data and determine the best fit.
Our ability to fit the data is limited by two types of noise: shot
noise and cosmic variance.  Shot noise arises because galaxy surveys
do not include an infinite number of objects, but it scales as the
inverse of the number of galaxies in the sample and is negligible when
$N \gg 1$.  The second source, cosmic variance, arises from the simple
fact that we live in only a particular realization of the ensemble of
possible universes.  No amount of galaxy sampling can eliminate this
noise, so it sets a theoretical upper limit to the accuracy of any
measurements that we might make.  In this section we will calculate
the intrinsic cosmic variance that we can expect in our measurements
of the moments of the power spectrum for a single window, representing
the ideal limit achievable by any redshift survey.

Let us begin with the usual definition for the covariance, namely 
$\sigma_{\ell,\ell'}^{2}(K,K') = \langle \bar{p}_{\ell'} 
(K')\bar{p}_{\ell} (K) \rangle - \langle \bar{p}_{\ell}(K) \rangle 
\langle \bar{p}_{\ell'} (K') \rangle$.  We choose different magnitudes 
for $K$ and $K'$ so as to treat the most general case.  Using our 
definition of $\bar{p}_{\ell}$ we can write
\begin{eqnarray}
\label{SIG1}
	\sigma_{\ell,\ell'}^{2}(K,K')& = & \left[\lfac \lpfac \int
	d\Omega_{K}d\Omega_{K'} {\cal P}(\mu_{K}) {\cal P}(\mu_{K'})
	{1 \over n^{4}V_{w}^{4}}\right.\\ \nonumber & & \times
	\sum_{i,j,k,l} {w(\vec{s}_{i})w(\vec{s}_{j})w(\vec{s}_{k})
	w(\vec{s}_{l}) \over
	\phi(\vec{s}_{i})\phi(\vec{s}_{j})\phi(\vec{s}_{k})\phi(\vec{s}_{l})}
	e^{i\vec{K}\cdot(\vec{s}_{i}-\vec{s}_{j})}
	e^{i\vec{K}'\cdot(\vec{s}_{k}-\vec{s}_{l})} \\ \nonumber & &
	\left. \times\left\langle (N_{i} - \langle N_{i} \rangle)
	(N_{j} - \langle N_{j} \rangle)(N_{k} - \langle N_{k}
	\rangle)(N_{l} - \langle N_{l} \rangle) \right\rangle
	\vphantom{{1 \over n^{4}V_{w}^{4}}}\right] \\ \nonumber &&-
	\langle \bar{p}_{\ell}(K)\rangle \langle \bar{p}_{\ell'}(K')
	\rangle.
\end{eqnarray}
In the limit that $n V_{w}$ grows large, {\it i.e} when ignoring shot
noise, it can easily be shown that the ensemble average of the product
of $N$'s is
\begin{equation}
\label{leading}
	 n^{4} d^{3}s_{i}~d^{3}s_{j}~d^{3}s_{k}~d^{3}s_{l}~
	\phi(\vec{s}_{i})\phi(\vec{s}_{j})\phi(\vec{s}_{k})\phi(\vec{s}_{l})
	\left[ \xi_{ij}\xi_{kl} + \xi_{ik}\xi_{jl} + \xi_{il}\xi_{kj}\right],
\end{equation} 
if the underlying distribution is Gaussian.  Restricting ourselves to
Gaussian fields is reasonable on large scales, particularly for
inflationary models of structure formation.  Putting things together,
we get
\begin{eqnarray}
\label{SIG2}
	 \sigma_{\ell,\ell'}^{2} & = & \left[ \lfac \lpfac \int
	 d\Omega_{K}d\Omega_{K'} {\cal P}(\mu_{K} ){\cal P}(\mu_{K'})
	 {1 \over V_{w}^{4}} \right. \\ \nonumber & & \times
	 \sum_{i,j,k,l} {w(\vec{s}_{i})w(\vec{s}_{j})w(\vec{s}_{k})
	 w(\vec{s}_{l}) } e^{i\vec{K}\cdot(\vec{s}_{i}-\vec{s}_{j})}
	 e^{i\vec{K}\cdot(\vec{s}_{k}-\vec{s}_{l})} \\ \nonumber &
	 &\left. \times \left[ \xi_{ij}\xi_{kl} + \xi_{ik}\xi_{jl} +
	 \xi_{il}\xi_{kj}\right]
	 d^{3}s_{i}d^{3}s_{j}d^{3}s_{k}d^{3}s_{l} \vphantom{{1 \over
	 V_{w}^{4}}}\right]\\ \nonumber && - \langle \bar{p}_{\ell}(K)
	 \rangle \langle \bar{p}_{\ell'}(K') \rangle
\end{eqnarray}
The three combinations of two-point correlation functions 
$\xi_{ij}\xi_{kl}$, $\xi_{ik}\xi_{jl}$, $\xi_{il}\xi_{kj}$, 
give us three different terms to calculate, which we call 
$I_{1}$, $I_{2}$, and $I_{3}$ respectively.  
Close inspection of $I_{1}$ reveals that it is 
exactly equivalent to $ \langle \bar{p}_{\ell}(K) \rangle  \langle 
\bar{p}_{\ell'}(K') \rangle$ canceling the final term.  
Thus we are left to calculate $I_{2}$ and $I_{3}$.  
 
Now let us take a closer look at $I_{2}$.  As before, we can do the 
spatial integrals and replace the window functions with their Fourier 
transforms.  The resulting equation is
\begin{eqnarray}
\label{I21}
	I_{2} & = & \lfac \lpfac \int d\Omega_{K}d\Omega_{K'} {\cal
	P}(\mu_{K} ){\cal P}(\mu_{K'}) \int {d^{3}k \over (2\pi^{3})}
	{d^{3}k' \over (2\pi^{3})} P(\vec{k}) P(\vec{k'}) \\ \nonumber
	& & \times \w^{(1)}(\vec{K}+\vec{k})
	\w^{(1)}(\vec{K}-\vec{k'})
	\w^{(1)}(\vec{K'}+\vec{k'})\w^{(1)}(\vec{K'}-\vec{k}).
\end{eqnarray}
We now expand the window function in a multipole series:
\begin{eqnarray*}
	 \w^{(1)}(\vec{K}+\vec{k})& 
		\equiv  &  \sum_{n} \w^{(1)}_{n}(k,K)
		{\cal P}_{n}(\mu_{k,K})  \\ \nonumber 
	 & = &  \sum_{n}\sum_{m=-n}^{n}\w^{(1)}_{n}(k,K) {4\pi \over 2n+1} 
	 Y_{n,m}(\Omega_{K}) Y_{n,m}^{*}(\Omega_{k}), 
\end{eqnarray*}
where $Y_{nm}$ is the usual spherical harmonic and the last line was 
derived from the addition theorem of spherical harmonics.  
Expanding similarly the other window functions, 
we may rewrite eq.\ (\ref{I21}) as
\begin{eqnarray}
\label{I22}
	I_{2} & = & \sum_{n_{1},m_{1}}\sum_{n_{2},m_{2}} 
	\sum_{n_{3},m_{3}}\sum_{n_{4},m_{4}} {(4 \pi)^{2}
	(2\ell+1)(2\ell'+1)	
	\over (2n_{1}+1)
	(2n_{2}+1)(2n_{3}+1)(2n_{4}+1)} 
	  \\ \nonumber 
	 &  & \times \int {d^{3}k \over (2\pi)^{3}} P(\vec{k}) 
	 (-1)^{n_{3}}\w^{(1)}_{n_{1}}(k,K)\w^{(1)}_{n_{3}}(k,K') 
		Y_{n_{1},m_{1}}^{*}
	 (\Omega_{k}) Y_{n_{3},m_{3}}^{*}(\Omega_{k})
	  \\ \nonumber 
	 &  & \times \int {d^{3} k'\over (2\pi)^{3}} P(\vec{k'}) 
	 (-1)^{n_{2}}\w^{(1)}_{n_{2}}(k',K)\w^{(1)}_{n_{4}}(k',K') 
		Y_{n_{2},m_{2}}^{*}
	 (\Omega_{k'}) Y_{n_{4},m_{4}}^{*}(\Omega_{k'})
	  \\ \nonumber
	 &  & \times \int d\Omega_{K} {\cal P}_{\ell}(\mu_{K}) 
	 Y_{n_{1},m_{1}}(\Omega_{K}) Y_{n_{2},m_{2}}(\Omega_{K})
	 \\ \nonumber 
	 & & 
	 \times \int d\Omega_{K'} {\cal P}_{\ell'}(\mu_{K'}) 
	 Y_{n_{3},m_{3}}(\Omega_{K'}) Y_{n_{4},m_{4}}(\Omega_{K'}).
\end{eqnarray}
Note that the factors $(-1)^{n_{2}}$ and $(-1)^{n_{3}}$ arise from the 
fact that ${\cal P}_{n}(-x) = (-1)^{n}{\cal P}_{n}(x)$. From the above
equation, 
we see two integrals of the form 
\begin{equation}
	\int d\Omega_{K} {\cal P}_{\ell}(\mu_{K}) 
	 Y_{n_{1},m_{1}}(\Omega_{K}) Y_{n_{2},m_{2}}(\Omega_{K}),
\end{equation}
which evaluate to  
\begin{equation}
	\sqrt{{4\pi \over 2\ell +1}} 
	C_{n_{1},\ell,n_{2};m_{1}}\delta_{m_{2},-m_{1}} ,
\end{equation}
where 
\( C_{n_{1},\ell,n_{2};m_{1}}\delta_{m_{2},-m_{1}}  = 0 \)  
unless the triangle condition \( |n_{1}-\ell| \leq n_{1} \leq 
n_{1}+\ell \) is satisfied, in which case 
\begin{equation}
	\label{CTERM}
	C_{n_{1},\ell,n_{2};m_{1}}\delta_{m_{2},-m_{1}}  = 
	\sqrt{{(2\ell+1)(2n_{1}+1)(2n_{2}+1)\over 4\pi}} \left(
	\begin{array}{ccc}
		n_{1} & \ell & n_{2}  \\
		0 & 0 & 0  \\
	\end{array}
	\right)
	\left(
	\begin{array}{ccc}
		n_{1} & \ell & n_{2}  \\
		m_{1} & 0 & -m_{1}
	\end{array}
	\right).
\end{equation}
The arrays in parentheses are the Wigner $3j$ symbols.
For a complete discussion of these symbols see, for example, 
\cite{edwards}.  Also, a fast recursive algorithm for calculating the $3j$ 
symbols is described in \cite{schulten}.  

The second type of integral which appears in eq.\ (\ref{I22}) has the 
form
\begin{equation}
	\int {k^{2}dk \over (2\pi)^{3}} \w^{(1)}_{n_{1}}(k,K) 
		\w^{(1)}_{n_{3}}(k,K') \int 
	d\Omega_{k} P(\vec{k}) Y_{n_{1},m_{1}}^{*}(\Omega_{k})
	Y_{n_{3},m_{3}}^{*}(\Omega_{k}).
\end{equation}
Expanding the power spectrum into its multipole moments, we can write 
this integral as 
\begin{equation}
	\sum_{i} {\cal S}_{n_{1},n_{3};i}(K,K') \sqrt{{4 \pi \over 2i+1}}
	 C_{n_{1},i,n_{3};-m_{1}} \delta_{m_{3},-m_{1}},
\end{equation}
where we have defined the function 
\begin{equation} 
\label{calS}
{\cal
 S}_{n_{1},n_{3};i}(K,K') \equiv \int {k^{2}dk \over (2\pi)^{3}}
 \w^{(1)}_{n_{1}}(k,K) \w^{(1)}_{n_{3}}(k,K') P_{i}(k).   \end{equation} 
Now we can combine all of the terms together to write 
\begin{eqnarray}
 I_{2} & = & \sum_{i,j}\sum_{n_{1},m_{1}}\ldots \sum_{n_{4},m_{4}}
 \delta_{m_{1},-m_{2}}\delta_{m_{1},-m_{3}}\delta_{m_{4},-m_{2}}
 \delta_{m_{4},-m_{3}} (-1)^{n_{2}+n_{3}}  \\ \nonumber & &
 \times {(4\pi)^{4}~{\cal S}_{n_{1},n_{3};i}(K,K') {\cal
 S}_{n_{2},n_{4};j}(K,K') \sqrt{2\ell+1} \sqrt{2\ell'+1}
\over
 (2n_{1}+1)(2n_{2}+1)(2n_{3}+1)(2n_{4}+1)
 \sqrt{2i+1}\sqrt{2j+1}}  \\
 \nonumber & & \times C_{n_{1},i,n_{3};-m_{1}}C_{n_{2},j,n_{4};-m_{1}}
 C_{n_{1},\ell,n_{2};m_{1}}C_{n_{3},\ell',n_{4};m_{3}}, 
\end{eqnarray} 
with the implicit constraints 
\begin{eqnarray*} 
  | n_{1} - \ell | & \leq n_{2} \leq & n_{1} + \ell, \\ | n_{1} - n_{3} |
  & \leq i \leq& n_{1}+n_{3}, \\ | n_{3} - \ell' | & \leq n_{4} \leq&
  n_{3} + \ell', \\ | n_{2} -n_{4}| & \leq j \leq& n_{2}+n_{4}.
\end{eqnarray*} 
Applying all of the constraints and summing out the
Kroniker delta's, $I_{2}$ takes the final form 
\begin{eqnarray} I_{2}
  & = & \sum_{n_{1},n_{3}} \sum_{n_{2}= |n_{1}-\ell|}^{n_{1}+\ell}
  \sum_{n_{4}=|n_{3}-\ell'|}^{n_{3}+\ell'} \sum_{i =
  |n_{1}-n_{3}|}^{n_{1}+n_{3}} \sum_{j = |n_{2}-n_{4}|}^{n_{2}+n_{4}}
  \sum_{m_1 = -n_1}^{n_1}
   \\ \nonumber & & \times (-1)^{n_{2}+n_{3}}{(4\pi)^{4} ~{\cal
  S}_{n_{1},n_{3};i}(K,K') {\cal S}_{n_{2},n_{4};j}(K,K') 
	\sqrt{2\ell+1}\sqrt{2\ell'+1}
	\over
  (2n_{1}+1)(2n_{2}+1)(2n_{3}+1)(2n_{4}+1)
  \sqrt{2i+1}\sqrt{2j+1}}  \\
  \nonumber & & \times C_{n_{1},i,n_{3};-m_{1}}C_{n_{2},j,n_{4};-m_{1}}
  C_{n_{1},\ell,n_{2};m_{1}}C_{n_{3},\ell',n_{4};m_{1}}.  
\end{eqnarray} 
A similar calculation can be performed for for $I_{3}$, but it should
be apparent from inspection that the result will be the same as for
$I_{2}$ except that there will no longer be the term
$(-1)^{n_{2}+n_{3}}$.  
Although calculating the infinite sum seems daunting, 
we found that, for the examples calculated in this paper, 
only a few moments in the power spectrum
and a few tens of moments in the window function 
were necessary to get very good convergence, 
making the problem numerically tractable.
 
\subsection{Shot Noise}
\label{shotNoise}
In the previous section we considered the variance in the multipole 
moments for a single measurement in the continuum limit.  For 
magnitude limited surveys, the number of galaxies contained in a 
given volume will fall with redshift until the variance arising from 
finite sampling, {\it i.e.}~shot noise, can become comparable and then 
dominate the cosmic variance.  To accurately determine the sensitivity 
of real surveys, one must accurately model the shot noise as well. In 
this sub--section, we shall discuss the shot noise calculation, 
leaving the details to the appendix.

Recall in eq.\ (\ref{SIG1}) that to calculate the covariance 
$\sigma^{2}_{\ell,\ell'}$ we needed to find the ensemble average of a 
four point moment
\begin{equation}
	\langle (N_{i} - \langle N_{i}\rangle ) (N_{j} - \langle N_{j}\rangle )
	(N_{k} - \langle N_{k}\rangle )(N_{l} - \langle N_{l}\rangle ) 
	\rangle,
	\label{4ptMoment}
\end{equation}
which was summed over all indices.  We claimed that in the limit of 
large $n$ this average could be reduced to eq.\ (\ref{leading}), and we 
will now show this explicitly while calculating all other terms.  We 
can rewrite the above average to explicitly include terms in which one 
or more of the indices are equal. The result is 
\begin{eqnarray}
	\label{allMomentTerms}
	 &&\langle (N_{i} - \langle N_{i}\rangle ) (N_{j} - \langle
	N_{j}\rangle )
	(N_{k} - \langle N_{k}\rangle )(N_{l} - \langle N_{l}\rangle ) 
	\rangle + 
	\\ \nonumber  
	&+&\bigl(\delta_{ij}\langle (N_{i} - \langle N_{i}\rangle )^{2}
	(N_{k} - \langle N_{k}\rangle )(N_{l} - \langle N_{l}\rangle ) 
	\rangle +  5~{\rm permutations} \bigr)
    \\	\nonumber
	 &+&\bigl(\delta_{ij}\delta_{kl}
	\langle (N_{i} - \langle N_{i}\rangle )^{2}
	        (N_{k} - \langle N_{k}\rangle )^{2}\rangle 
	  +  2~{\rm permutations}  \bigr)
	\\ \nonumber  
	&+&\bigl(\delta_{ij}\delta_{jk}
	\langle (N_{i} - \langle N_{i}\rangle )^{3}
		(N_{l} - \langle N_{l}\rangle )\rangle  
	  +  3~{\rm permutations} \bigr)
    \\ \nonumber 
	&+&\bigl(\delta_{ij}\delta_{jk}\delta_{kl} 
	\langle (N_{i} - \langle N_{i}\rangle )^{4}\rangle,  
\end{eqnarray}
where different indices are presumed to be unequal. Since
in the limit of infinitesimal volumes $\langle N_{i}^{n} \rangle = 
n\phi(\vec{s}_{i})d^{3}s_{i}$, the ensemble averages reduce to 
\begin{eqnarray}
  	\label{evaluatedTerms}
	 &  & n^{4}\phi(\vec{s}_{i})\phi(\vec{s}_{j})\phi(\vec{s}_{k})
 	 \phi(\vec{s}_{l})d^{3}s_{i}d^{3}s_{j}d^{3}s_{k}d^{3}s_{l}
 	 \left[\xi_{ij}\xi_{kl}+\xi_{ik}\xi_{jl}+\xi_{il}\xi_{jk}\right]
 	\\ \nonumber 
 	 & + & \bigl( \delta_{ij} n^{3}\phi(\vec{s}_{j})\phi(\vec{s}_{k})
 	 \phi(\vec{s}_{l})d^{3}s_{j}d^{3}s_{k}d^{3}s_{l}\xi_{kl} + 
 	 5~{\rm permutations} \bigr)
 	  \\ \nonumber 
 	 & + & \bigl( \delta_{ij}\delta_{kl} n^{2} \phi(\vec{s}_{i})\phi(\vec{s}_{k})
 	 d^{3}s_{i}d^{3}s_{k}(1+\xi_{ik}) + 2~{\rm permutations}\bigr)
 	  \\ \nonumber
 	 & + & \bigl( \delta_{ij}\delta_{jk}\phi(\vec{s}_{k})\phi(\vec{s}_{l})
 	  d^{3}s_{k}d^{3}s_{l}\xi_{kl} +  3~{\rm permutations} \bigr)
 	  \\ \nonumber 
 	 & + & \delta_{ij}\delta_{jk}\delta_{kl} n \phi(\vec{s}_{k}) 
	d^{3}s_{i},
\end{eqnarray}
for Gaussian fields.  The highest order term in $n$ is the one which 
we have previously calculated, while the others are small in 
comparison when $n$ is large.  What remains is to substitute eq.\ 
(\ref{evaluatedTerms}) into eq.\ (\ref{SIG1}) and evaluate all the 
terms.  These details are left to the appendix, but here we give the 
results.  To all orders in $n$, the shot noise contribution to the 
variance of the multipole moments of the power spectrum is
\begin{eqnarray}
\label{shotnoise}
	\sigma_{shot~\ell,\ell'}^2(K,K') 
	 & = & {\sqrt{(2\ell+1)(2\ell'+1)}} {1 \over N} 
	\sum_{i,j,m_{j}}\sum_{n_{1}= |\ell -i|}^{\ell +i}
	\sum_{n_{2}= |\ell' -j|}^{\ell' +j}\Biggl\{ 
	\biggl[ 2\left[ (-1)^{n_{2}+m_{j}} 
	(-1)^{j+m_{j}} \right] \\ \nonumber
	&\times& { (4\pi)^{5/2}
	\over \sqrt{2i+1}(2n_{1}+1)(2n_{2}+1)(2j+1)}
	 \tilde{w}^{(2)}_{j}(K,K') \\ \nonumber
	&\times&
	{\cal S}_{n_{1},n_{2};i} 
	 C_{n_{1},i,n_{2};m_{j}} C_{n_{1},\ell,i;-m_{j}} C_{n_{2},\ell'
	,n_{2};m_{j}}\biggr] \\ \nonumber
	&+& 
	 \delta_{\ell 0}\delta_{\ell' 0}   {4\pi \over N^2} 
	\int {k^2 dk \over (2\pi)^3}  \tilde{W}^{(2,2)}(k) P_0(k) 
	+
	\delta_{\ell \ell'}\left[ 1 + (-1)^\ell \right]
	 {1 \over N^2} \tilde{W}^{(2,2)}(K,K') \\
	\nonumber 
	&+& 
	\Biggl[ \left( {4\pi \over N} \right)^2 {\bar{V} \over V_w}
	(2 \ell +1)(2\ell'+1)
	\sum_n \int{k^2 dk \over (2\pi)^3} P_n(k) \int {s^2ds\over
	V_w} G(s) \left( i^{\ell+\ell'+n} + i^{\ell-\ell'+n} \right) 
	\\ \nonumber
	&\times& j_\ell(Ks) j_{\ell'}(K's) j_n(ks)	
	\left(
	\begin{array}{ccc}
		\ell & \ell' & n  \\
		0 & 0 & 0
	\end{array}
	\right)^2\Biggr]
	\\ \nonumber 
	&+&  
	 \delta_{\ell 0}{2 \over N^2} {4\pi \over 2\ell'+1}
	\int{k^2 dk \over (2\pi)^3}\tilde{W}^{(3,1)}_{\ell'}(K',k) 
	P_{\ell'}(k) \\ \nonumber
	&+& \delta_{\ell' 0} {2 \over N^2} {4\pi \over 2\ell+1}
	\int{k^2 dk \over (2\pi)^3}\tilde{W}^{(3,1)}_\ell(K,k) P_\ell(k)
	\\ \nonumber
	&+&  \delta_{\ell 0} \delta_{\ell' 0}{1 \over N^3} 
	\int {d^3 s \over V_w} w^4(s)\Biggr\},
\label{sigmaShot}	
\end{eqnarray}
where 
\begin{equation}
	G(\vec{s})  \equiv {1\over 2} \int {d^3 s'\over V_w^2} 
		w^2({\vec{s}+\vec{s'}\over 2})
		w^2({\vec{s}-\vec{s'}\over 2})
\end{equation}
and $\tilde{w}^{(p)}(\vec{k})$, $\tilde{W}^{(p,q)}(\vec{k})$ and 
$\tilde{W}^{(p,q)}_\ell(\vec{k})$ are defined above.  In calculating 
these terms, we have assumed that the window function is spherically 
symmetric (implying $G(\vec{s})=G(s)$), and also that the selection 
function $\phi$ is constant over the window, leading us to define $N = 
n \phi V_w$.

Since factors of $n$ now appear explicitly in the shot noise terms, we
require an estimate of the number density of galaxies as a function of
redshift.  The SDSS is an $r$-band magnitude-limited survey, so we
need an appropriate galaxy luminosity function.  An analysis of the
same band for the galaxies in the Las Campanas redshift survey
performed by \cite{lin} has shown that the galaxy luminosity function
is well fit by a Schechter function with $M^* = -20.29 \pm 0.02
+5~\log h$, $\alpha = -0.70 \pm 0.05$, and $\phi^* = 0.019 \pm
0.001~h^3$.  For our models we will presume that these parameters will
also well describe the SDSS and take $h= 0.7$, consistent with the
fiducial models we shall consider.  To get $10^6$ galaxies in the
total survey volume, we choose an apparent magnitude limit of $m =
17.6$ consistent with the limit $m \approx 18$ quoted for the SDSS.  A
second map has also been proposed for the SDSS data that would generate a
volume limited survey of $10^5$ bright red galaxies with an
approximate depth of $z= 0.5$ (\cite{szalay}). We found in our
calculations that the density of galaxies for this survey was
insufficient to significantly strengthen the constraints derived from
the proposed magnitude limited survey, so we shall only analyze the
latter in detail.
 
\section{Sensitivity of Redshift Surveys} 
The next generation of redshift surveys, particularly the Sloan
Digital Sky Survey, will probe large fractions of the sky observing up
to a million galaxy redshifts.  In this section, we would like to
estimate how well SDSS, the largest proposed survey, can measure
cosmological parameters from redshift space distortions.
Specifically, we want to determine whether the geometric and
evolutionary effects which occur at larger redshifts can break the
degeneracy between the matter density and the bias, producing a clean
signal for $\Omega_0$.  To that end, we consider a pair of fiducial
models which have good concordance with observation and perform a
statistical fit of cosmological parameters.

In the last section, we calculated the mean and variance of the
multipole moments of the power spectrum measured in small sub--volumes
of a large survey.  For nearby samples, when geometric effects are
unimportant, the ratios of any two moments depend only on $\beta$ and
the shape of the power spectrum, but not its normalization.
Cosmological models are roughly degenerate in parameter choices for
which $\beta = \Omega_0^{0.6}/b_0$ is a constant.  We shall reconsider
these ratios with the inclusion of geometric and evolutionary effects
to see if this degeneracy can be broken in deeper surveys.  
The ratio of $\bar{p}_2/\bar{p}_0$ is considered as it yields the largest
signal to noise proportion; our numerical results show that the non--linear
clustering effects tend to suppress the signal in higher moments. Two
types of fiducial models with $\Omega_0 = 0.3$ are considered, one an
open model with $\lambda = 0$ and the other a flat model with $\lambda
= 0.7$, as these choices are favored by observations of large scale
structure, {\it e.g}.\ \cite{tadros}. For the dark matter component,
we assume Cold Dark Matter, with a fixed shape parameter of $\Gamma
\equiv \Omega_0 h_0 = 0.2$, which we use for all test
models so as to deconvolve the redshift distortion effects from the
effects of the of the changing power spectrum shape.  Furthermore,
$\Gamma$ should be well measured in the upcoming redshift surveys by
direct calculation of the power spectrum, and thus should be a fixed
input.  To model the non--linear dispersion, we choose a velocity of
300 km/s which corresponds to a length scale of 3 $h^{-1}$ Mpc in
Hubble units.

In comparing our test models to the fiducial ones, we must consider 
how one can best extract measurements of $\bar{p}_2/\bar{p}_0$ from a 
real survey.  \cite{cole} examined n-body simulations in the case of 
no evolution or geometric effects.  They took repeated sub--samples by 
randomly locating the center of a window function and measuring the 
multipole moments in these sub--volumes.  To ensure that the data are 
statistically independent, the typical separation distance between 
sub--samples is chosen to be the size of the window, otherwise, the 
overlap will introduce correlations.  We reconsider this procedure, 
only  now, the effects of redshift evolution and geometry will be 
included to more accurately model real surveys.

Clearly, to maximize the number of measurements, one ought to consider
the smallest possible sub--volume.  Unfortunately, non--linear effects
dominate on small scales and our simple model of the non--linear
velocity dispersion will break down.  In n-body simulations,
\cite{cole} found that the corrected linear models failed at
wavelengths below about 20$h^{-1}$ Mpc, so we shall take this to be
the lower limit on which we can apply our linear calculations.  Thus
we consider a Gaussian window---chosen because its multipole moments
are expressible analytically---with a radius in redshift space which
corresponds locally to $r_0 = 20h^{-1}$ Mpc, where $r = H_0^{-1} z$.
For our calculations of $\bar{p}_\ell(K)$, we selected $K =
2\pi/20~h{\rm Mpc}^{-1}$ consistent with the n-body results.  We then
divided space into slices $\pi z^2dz$ corresponding to the $\pi$
steradians to be covered by the SDSS, with $dz$ equal to the window
diameter of $40 h^{-1}$ Mpc locally.  Redshifts corresponding to less
than twice the window width were ignored because they were not
sufficiently distant for the distant observer ({\it i.e.} small angle)
approximation assumed in eq.\ (\ref{xofz}) of \S \ref{distortion} to
be valid.  High redshift data was dropped once it became shot-noise
dominated and no longer contributed significantly to the fits.  In
each of these slices we divided the shell volume by the window volume
to determine the number of statistically independent measures that
were available at that redshift.  To determine the expected variance
of the $\bar{p}_\ell(K)$'s, we divided the variance for a single
measurements by the number of volumes, the standard suppression for
multiple independent measurements, and calculated the errors in their
ratios using simple propagation of errors.  In other words, the error
bars were determined by %
\begin{equation}
\sigma^2 = {1 \over {\cal N}}{\p_2^2 \over \p_0^2}\left( {\sigma^2_{00} \over
\p_0^2} -2{\sigma^2_{02} \over \p_0 \p_2} + {\sigma^2_{22} \over
\p_2^2}\right),
\end{equation}
where ${\cal N}$ is the number of independent volumes in the particular
redshift slice.  The result was a set of data points showing the
expected mean and deviation for the redshift bins one would reasonably
choose when analyzing a real survey.  We then repeated this 
procedure for $r_{0} = 40h^{-1}$ Mpc, the smallest scale which is 
independent of the $20h^{-1}$ Mpc data.  We found that data on both 
scales were necessary to constrain all the model parameters, while 
including larger scales was not significantly more constraining.  

To compare with other cosmologies, we calculated the mean values of
$\bar{p}_2/\bar{p}_0$ for models with varying $\Omega_{0}$, $b_{0}$,
and $\sigma$ and calculated a $\chi^{2}$ using the fiducial model. We
defined our confidence limit as the surface of constant $\chi^{2}$
into which a given percentage of best fit model parameters would fall
for multiple realizations of the fiducial ensemble.  For example, the
95\% confidence limit is defined such that the best fit model
parameters will fall within that surface 95\% of the time. Since we
have three fitting parameters, there are three degrees of freedom in
our $\chi^2$ statistics.

\section{Results and Conclusions}
In figure \ref{compare}, we show two examples of $\bar{p}_2/\bar{p}_0$
plotted as a function of redshift, where we have removed the shot
noise contributions to the mean values.  In the upper panel, we show a
fiducial model (stars) of $\Omega_0 = 0.3$, $\lambda_0 = 0.7$ and $b_0
= 1.0$ with $1\sigma$ error bars, compared to a test model (boxes) of
$\Omega_0 = 0.48$, $\lambda_0 = 0.52$, and $b_0 = 1.2$; and in the
lower panel, we show a fiducial model of $\Omega_0 = 0.3$, $\lambda_0
= 0$, and $b_0 = 1.0$ compared to a test model of $\Omega_0 = 0.85$,
$\lambda_0 = 0$, and $b_0 = 1.60$.  In each example, data for both the
$20 h^{-1}$ Mpc and $40 h^{-1}$ Mpc windows are shown, where the
larger window produces higher values for the ratio, and all models use
the fiducial value for $\sigma$.  Data was terminated at the redshift
after which shot noise dominated the result.  Both comparison models
are barely accepted at the 95\% confidence limit when the fiducial
velocity dispersion was used for each.  We emphasize this with figure
\ref{O.3L.7flat} which shows the 68\%, 95\%, and 99\% confidence
limits for flat, cosmological constant test models given a fiducial
$\Omega_0 = 0.3$, $\lambda_0 = 0.7$, and $b_0 = 1.0$ model like that
of figure \ref{compare}.  These surfaces are projections of the full
three dimensional confidence volume, and they represent the confidence
limits when the non--linear velocity dispersion is unconstrained by
other data.  The
dotted line shows the curve $\Omega_0^{0.6}/b_0 = 0.49$ which is the
degeneracy one expects locally when redshift effects are ignored.  We
do not show a figure comparing open models to our fiducial
cosmological constant model, as none were acceptable at even the 99\%
confidence limit.  In figure \ref{O.3L0curve} we switch to an open
$\Omega_0 = 0.3$, $\lambda_0 = 0$, and $b_0 = 1.0$ fiducial model and
test it against other open models.  Finally, in figure \ref{O.3L0flat}
we test this fiducial model against flat cosmological constant models.

In figure \ref{compare} we see the significant redshift dependence of
$\bar{p}_2/\bar{p}_0$, where we would expect none in models which
ignore geometry and evolution.  Naively fitting a horizontal line
representing models with no redshift dependent effects, we get a best
fit $\beta_{0} = 0.45$ for our fiducial flat model and $\beta_{0} =
0.26$ for our fiducial open model
using only the $20 h^{-1}$ Mpc data; we expect a value of $\beta_{0}
\approx 0.5$ in both cases.  Thus we verify that there will be
systematic errors in models which ignore redshift evolution effects, a
point emphasized by \cite{nakamura}.  The slopes of theses data can be
qualitatively understood by considering the competing effects of the
evolution of $\beta$ and geometry.  At deeper redshifts, $\Omega(z)
\rightarrow 1$ and $\beta$ increases, tending to increase the value
of the ratio of $\bar{p}_2/\bar{p}_0$ for the power spectrum and
non--linear dispersion we have chosen.  Geometry, on the other hand,
pushes the effective scale which is probed to smaller scales and thus
larger values of the wave number, {\it i.e}.\ objects appear more
spread out at high redshift.  Non--linear dispersion tends to drive
the multipole ratio to lower values as the effective $K$ grows, so
geometric effects tend to decrease the multipole ratios.  Thus we have
two competing effects which determine the slope of the multipole ratio
curve as a function of redshift.  For $\Omega_0 = 1$ models, $\beta$
remains fairly constant---changing only due to the evolution in bias.
These models are dominated by geometric effects and thus have the
steepest negative slope.  The high cosmological constant model has the
flattest slope, because the rapid evolution in $\beta$ just cancels
the geometric effects, while the open models fall in between. The
effects of changing the non--linear velocity dispersion generally
shifts the overall normalization of the data without affecting the
slope significantly.  With only the $20 h^{-1}$ Mpc data, there is a
degeneracy between $b_0$ and $\sigma$.  To break it, we need to
include the data from the larger window, as the change in
normalization is different for the two window scales.  We also note
that the nearly flat evolution of the multipole ratio also explains
why open models cannot fit our fiducial cosmological model; all are
too sloped to be good fits to the data.

Regarding the determination of cosmological parameters, the results
shown in the last three figures reveal interesting results.  For the
fiducial open model (figure \ref{O.3L0curve}), the degeneracy between
models of differing $\Omega_0$ and $b_0$ is weakly broken, and the
redshift distortions do permit us to distinguish between the open
fiducial model and $\Omega_0 = 1$ models at the 95\% limit, although
not at the 99\% limit.  However, when we tried a fiducial $\Omega_0
=0.4$ model, $\Omega_0 = 1$ was allowed at the 95\% limit, so one can
conclude that an open universe with $\Omega_0 = 0.3$ is just on the
edge of being able to exclude critical matter density models.  For the
case of the flat fiducial model (figure \ref{O.3L.7flat}), we see that
values of $\Omega_{0} > 0.48$ are ruled out at the 95\% confidence
limit, producing significantly stronger constraints than is the case
for the open model.  In figure
\ref{O.3L0flat} we test the likelihood of confusing flat models with
open.  We see that models with some cosmological constant can be
confused with the open fiducial model.  Overall, we conclude that
redshift surveys like SDSS may just be able to determine bias and
evolution independent measures of cosmological parameters, at least in
discriminating the extremes of $\Omega_0 = 1$ and $\Omega_0 =
0.3$. although only nominally so for open universes.
We also see that naive estimates of $\beta$ which ignore redshift
evolution are systematically biased towards smaller values.

When it comes to determining cosmological parameters, the ideal result
is to measure them with several independent observations, looking for
a consistency of result which would demonstrate that we understand the
fundamentals of cosmology or an inconsistency that would indicate a
failure of some aspect of current theory.  It has been suggested that
redshift surveys offer promise of being a reliable source of data for
such cosmological parameter determination, when appropriate statistics
are applied.  Simply measuring the change in the mean number of
galaxies as a function of redshift is inadequate, because we cannot
observe the gravitational component in the absence of reliable models
for the non--gravitational evolution.  Alternatively, one can consider
the redshift evolution of the multipole moments of the linear power
spectrum, which are driven only by gravity, and hope to cleanly
observe cosmological effects.  
Our results suggest that 
if the true model of the universe contains a large cosmological constant,
then $\Omega_0$ is tightly constrained; however,
for open models
the the extremes of $\Omega_0 = 1$ and $\Omega_0 = 0.3$ only 
can marginally be distinguished (95\% but not 99\% confidence limit)
in surveys on the scale of SDSS.

\newpage 
\appendix
\begin{center}Appendix A. \end{center}
In \S\ref{shotNoise}, we discussed the complete expansion of the 
variance including the lower order terms in $n$.  Here we shall show 
the details of calculating the shot noise terms ignored in 
\S\ref{covar}. We begin with the leading order correction in eq.\ 
(\ref{allMomentTerms}), which when substituted into eq.\ (\ref{SIG1}), 
may be written
\begin{eqnarray}
\label{order1overN}
	& & \lfac \lpfac \int
	d\Omega_{K}d\Omega_{K'} {\cal P}(\mu_{K}) {\cal P}(\mu_{K'})
	{1 \over n^{4}V_{w}^{4}}\\ \nonumber & \times& 
	\sum_{i,j,k,l} {w(\vec{s}_{i})w(\vec{s}_{j})w(\vec{s}_{k})
	w(\vec{s}_{l}) \over
	\phi(\vec{s}_{i})\phi(\vec{s}_{j})\phi(\vec{s}_{k})\phi(\vec{s}_{l})}
	e^{i\vec{K}\cdot(\vec{s}_{i}-\vec{s}_{j})}
	e^{i\vec{K}'\cdot(\vec{s}_{k}-\vec{s}_{l})} \\ \nonumber &\times &
	 \left[
	 \delta_{ij}\langle (N_{i} - \langle N_{i} \rangle)^{2}
	(N_{k} - \langle N_{k}\rangle)(N_{l} - \langle N_{l} \rangle) 
	\rangle 
	+ \delta_{ik}\langle (N_{i} - \langle N_{i} \rangle)^{2}
	(N_{j} - \langle N_{j}\rangle)(N_{l} - \langle N_{l} \rangle) 
	\rangle \right.
	\\ \nonumber  &+&
	  \delta_{il}\langle (N_{i} - \langle N_{i} \rangle)^{2}
	(N_{j} - \langle N_{j}\rangle)(N_{k} - \langle N_{k} \rangle) 
	\rangle 
	+ \delta_{jk}\langle (N_{i} - \langle N_{i} \rangle)
	(N_{j} - \langle N_{j}\rangle)^{2}(N_{l} - \langle N_{l} \rangle) 
	\rangle 
	\\ \nonumber &+& \left.
	  \delta_{jl}\langle (N_{i} - \langle N_{i} \rangle)
	(N_{j} - \langle N_{j}\rangle)^{2}(N_{k} - \langle N_{k} \rangle) 
	\rangle 
	+ \delta_{kl}\langle (N_{i} - \langle N_{i} \rangle)
	(N_{j} - \langle N_{j}\rangle)(N_{l} - \langle N_{l} \rangle)^{2}
	\rangle \right].
\end{eqnarray}
Now let us digest this term by term.  The first we shall designate 
$I_{1,1}$ and, recalling eq.\ (\ref{evaluatedTerms}), we may write it 
as
\begin{eqnarray}
	I_{1,1} & =  & \pllint {1 \over n V_{w}^{4}} \sum_{j,k,l} 
	{w^{2}(\vec{s}_{j}) \over \phi(\vec{s}_{j})}w(\vec{s}_{k})
	w(\vec{s}_{l})
	 \\ \nonumber 
	 & \times & \xi_{k,l} e^{i\vec{K'}\cdot (\vec{s}_{k}-\vec{s}_{l})}
	 d^{3}s_{j}d^{3}s_{k}d^{3}s_{l}.
	\label{I11,1}
\end{eqnarray}
To allow us to evaluate many of the integrals that we will encounter
in this section, we need to make a simplifying assumption about the
selection function $\phi$: that it is approximately constant in the
window of interest.  For small windows this should be reasonable, so
we thus define $N = n\phi V_{w}$ to be the number of galaxies in a
given window. Substituting in the Fourier expansion for the
correlation function and evaluating the spatial integrals, it is
straight forward to show
\begin{equation}
	I_{1,1}  =  \delta_{l0}{1 \over N}{\bar{V} \over V_{w}}
	\int d \Omega_{K'} {\cal P}_{\ell'}(\mu_{K'}) \int {d^{3}k \over 
	(2\pi)^{3}} \tilde{W}^{(1,1)}(\vec{K}+\vec{k'}) P(\vec{k}),
	\label{I11,2} 
\end{equation}
with 
\begin{equation}
	\bar{V} = \int d^{3}s~w^{2}(\vec{s}).
	\label{barV}
\end{equation}
and $W^{(p,q)}$ defined in eq.\ (\ref{Wpq}).
Expanding everything into Legendre series and evaluating the angular 
integrals, one can show directly that
\begin{equation}
\label{I1_1final}
	I_{1,1} = \delta_{l0} {4\pi \over 2 \ell' + 1} {\bar{V} \over V_{w}}
	\int {k^{2}dk \over (2\pi)^{3}} P_{\ell'}(k) 
	\tilde{W}^{(1,1)}_{\ell'}(K',k).
	\label{I11final}
\end{equation}
By inspection of eq.\ (\ref{order1overN}), one can also see that 
$I_{1,6}$ is equal to $I_{1,1}$ under the interchange of $\ell$ and 
$\ell'$.  However, looking at eq.\ (\ref{PBfinal}),
we see that both of these terms will cancel with the terms 
coming from $\langle \bar{p}_{\ell} \rangle \langle \bar{p}_{\ell'} 
\rangle$, so they are dropped below.  

Now we continue on to consider the second term $I_{1,2}$ which may be 
written  
\begin{eqnarray}
	I_{1,2} & =  & \pllint {1 \over n V_{w}^{4}} \sum_{j,k,l} 
	{w^{2}(\vec{s}_{k}) \over \phi(\vec{s}_{k})}w(\vec{s}_{j})
	w(\vec{s}_{l})
	 \\ \nonumber 
	 & \times & \xi_{j,l} e^{i\vec{K}\cdot (\vec{s}_{k}-\vec{s}_{j})}
	 e^{i\vec{K'}\cdot (\vec{s}_{k}-\vec{s}_{l})}
	 d^{3}s_{j}d^{3}s_{k}d^{3}s_{l}.
	\label{I12,1}
\end{eqnarray}
Performing the spatial integrals, we find
\begin{eqnarray}
	I_{{1,2}} &=& {1 \over N} \pllint \\ \nonumber
	&& \int {d^{3}k \over (2\pi)^{3}} \tilde{w}^{(2)}(\vec{K} + \vec{K'})
	\tilde{w}(\vec{k} + \vec{K}) \tilde{w}(\vec{k} - \vec{K'}) P(\vec{k}).
	\label{I12,2}
\end{eqnarray}
with $\tilde{w}^{(2)}$ defined in eq.\ (\ref{wp}).
Let us take a closer look at the integral 
\begin{equation}
	\int {d^{3}k \over (2\pi)^{3}} P(\vec{k}) \tilde{w}(\vec{k}+\vec{K})
	\tilde{w}(\vec{k}-\vec{K'}).
\end{equation}
If we expand each term into a Legendre series, and then expand those
Legendre polynomials into spherical harmonics using the addition
theorem, we can see after evaluating the angular integrals that this
term may be reduced to
\begin{equation}
	\sum_{n_{1},n_{2}} \sum_{i= 
	|n_{1}-n_{2}|}^{n_{1}+n_{2}}\sum_{m_{1}=-n_{1}}^{n_{1}} {(4 \pi)^{2} 
	\over (2 n_{1} +1) (2 n_{2} +1)} {\cal S}_{n_{1},n_{2};i} (-1)^{n_{2}} 
	\sqrt{{4\pi \over 2 i+ 1}} Y_{n_{1},-m_{1}}(\Omega_{K})
	Y_{n_{2},m_{1}}(\Omega_{K'}) C_{n_{1},i,n_{2};m_{1}}
\end{equation}
with ${\cal S}$ defined in eq.\ (\ref{calS}).
To make further progress, we substitute the above back into eq.\ 
(\ref{I12,2}) and expand $\tilde{w}^{(2)}(\vec{K} + \vec{K'})$ in a 
spherical harmonic series. The rest of the work consists in 
evaluating various angular integral like we have already seen, so we 
shall just quote the final result:
\begin{eqnarray}
\label{I1_2final}
	I_{1,2} & = & {\sqrt{(2\ell+1)(2\ell'+1)}} {1 \over N} 
	\sum_{i,j,m_{j}}\sum_{n_{1}= |\ell -i|}^{\ell +i}
	\sum_{n_{2}= |\ell' -j|}^{\ell' +j} (-1)^{n_{2}+m_{j}} \\ \nonumber 
	&\times& { (4\pi)^{5/2}
	\over \sqrt{2i+1}(2n_{1}+1)(2n_{2}+1)(2j+1)}
	  \\ \nonumber 
	 & \times & \tilde{w}^{(2)}_{j}(K,K') {\cal S}_{n_{1},n_{2};i} 
	 C_{n_{1},i,n_{2};m_{j}} C_{n_{1},\ell,i;-m_{j}} C_{n_{2},\ell'
	,n_{2};m_{j}}. 
\end{eqnarray}
By inspection one can see that the term $I_{1,5}$ is equal to 
$I_{1,1}$.  The remaining terms $I_{1,3}$ and $I_{1,4}$ are equivalent 
to $I_{1,2}$ under the transformation $-1^{n_{2}} \rightarrow -1^{j}$.

The next correction term in eq.\ (\ref{allMomentTerms}) has the form:
\begin{eqnarray}
\label{order1overN2_1}
	& & \lfac \lpfac \int
	d\Omega_{K}d\Omega_{K'} {\cal P}(\mu_{K}) {\cal P}(\mu_{K'})
	{1 \over n^{4}V_{w}^{4}}\\ \nonumber & \times& 
	\sum_{i,j,k,l} {w(\vec{s}_{i})w(\vec{s}_{j})w(\vec{s}_{k})
	w(\vec{s}_{l}) \over
	\phi(\vec{s}_{i})\phi(\vec{s}_{j})\phi(\vec{s}_{k})\phi(\vec{s}_{l})}
	e^{i\vec{K}\cdot(\vec{s}_{i}-\vec{s}_{j})}
	e^{i\vec{K}'\cdot(\vec{s}_{k}-\vec{s}_{l})} \\ \nonumber &\times &
	 \left[
	 \delta_{ij}\delta_{kl}\langle N_j N_l \rangle  
	 + 
	 \delta_{ik}\delta_{jl}\langle N_k N_l \rangle 
	 + 
	 \delta_{il}\delta_{jk}\langle N_l N_k \rangle 
	 \right].
\end{eqnarray}
We refer to each of the terms here as $I_{2,1},~I_{2,2}$ and $I_{2,3}$
respectively. The first, $I_{2,1}$ can be expanded into the
following:
\begin{eqnarray}
\label{I2_1}
	I_{2,1} &=& \delta_{\ell 0}\delta_{\ell' 0} {1 \over N^2}
	\left[ \left({\bar{V} \over V_w} \right)^2 + \int {d^3k \over
	(2\pi)^3} \tilde{W}^{1,1} P(\vec{k}) \right],
\end{eqnarray}
after we insert the Fourier expansion for the correlation function and
evaluate the spatial integrals.  The first term will cancel with
pieces from $\langle \bar{p}_{\ell} \rangle \langle \bar{p}_{\ell'}
\rangle$, leaving the second.  The angular part of the $k$ integral is
easily evaluated, leaving
\begin{equation}
\label{I2_1final}
	I_{2,1} = \delta_{\ell 0}\delta_{\ell' 0} {4 \pi \over N^2}
	\int{k^2 dk \over (2\pi)^3} \tilde{W}^{(1,1)}(k) P_0(k).
\end{equation}

For the second term in eq.\ (\ref{order1overN2_1}),  we shall not
automatically evaluate the spatial integral because the results are
simpler to reduce for some terms if we evaluate the angular parts
first. Thus we write
\begin{eqnarray}
\label{I22_1}
	I_{2,2} &=& {1 \over N^2} {2\ell +1 \over 4\pi}{2\ell' +1 \over 4\pi}  
	\int d\Omega_K d\Omega_K' {\cal P}_l(\mu_K){\cal P}_l'(\mu_K')
	\sum_{k,l} w^2(\vec{s}_k)w^2(\vec{s}_l) \\ \nonumber 
	& \times & e^{i \vec{s}_k \cdot (\vec{K} +\vec{K}')}
	e^{-i \vec{s}_l \cdot (\vec{K} +\vec{K}')}
	\left[1 + \int{d^3k \over (2\pi)^3}
	e^{i\vec{k}\cdot(\vec{s}_k-\vec{s}_l)}P(\vec{k}) \right] 
	d^3s_k d^3s_l.
\end{eqnarray}
The first term, designated $I_{2,2A}$ evaluates to 
\begin{equation}
I_{2,2A} = {1\over N^2}\pllint \tilde{W}^{(1,1)}(\vec{K} +\vec{K'}),
\end{equation}
which, after expanding the widow function into a Legendre series and
evaluating the angular integrals, reduces to 
 \begin{equation}
\label{I2_2Afinal}
I_{2,2A} = \delta_{\ell \ell'}{1\over N^2} \tilde{W}^{(1,1)}_\ell(K,K').
\end{equation}

To tackle the second term in eq.\ (\ref{I22_1}), we first define a new
set of spatial variable $\vec{s} = \vec{s}_k - \vec{s}_l$ and 
$\vec{s}' = \vec{s}_k-\vec{s}_l$. We rewrite the spatial sums as integrals in
the new variables, producing
\begin{eqnarray}
	I_{2,2B} &=& {1\over N^2} \pllint   \\ \nonumber
	&\times& 
	{1\over 2} \int{d^3 s d^3 s' \over V_w^2}
	w^2({\vec{s}+\vec{s}'\over 2}) w^2({\vec{s}-\vec{s}'\over 2}) 
	e^{iK\cdot\vec{s}} e^{iK'\cdot\vec{s}} 
	\int {d^3 k \over (2\pi)^3} 
	e^{i\vec{k}\cdot\vec{s}}P(\vec{k}).
\end{eqnarray}
The quantity
\begin{equation}
G(\vec{s}) \equiv 
	{1\over 2} \int{d^3 s d^3 s' \over V_w^2}
	w^2({\vec{s}+\vec{s}'\over 2}) w^2({\vec{s}-\vec{s}'\over 2}) 
\end{equation}
is a function only of the magnitude of $\vec{s}$ if $w$ is spherically
symmetric.
The exponential has the following expansion in Legendre polynomials
\begin{equation}
e^{iK\cdot\vec{s}} = \sum i^n (2n+1) j_n(Ks){\cal P}_n(\mu_{K,s}),
\end{equation}
which we may use, along with eq.\ (\ref{LL}) and eq.\ (\ref{CTERM})
to evaluate the angular integrals. The final result can be written
\begin{eqnarray}
\label{I2_2Bfinal}
	I_{2,2B} &=& \left({4\pi \over N} \right)^2 (2\ell+1)(2\ell'+1)
\sum_n \int {k^2 dk \over (2\pi)^3} P_n(k) \int{s^2 ds\over V_w} G(s)
\\ \nonumber
	&\times&
	\left(
	\begin{array}{ccc}
		\ell & \ell' & n  \\
		0 & 0 & 0
	\end{array}
	\right)^2
	i^{\ell+\ell'+n} j_\ell(Ks) j_{\ell'}(K's) j_n(ks).	
\end{eqnarray}
Inspection of the third term in eq.\ (\ref{I22_1}) reveals that
$I_{2,3A} = (-1)^\ell I_{2,2A}$ while $I_{2,3B} = i^{-2\ell'}I_{2,2B}$.

Moving along, we see that the fourth term in eq.\
(\ref{allMomentTerms}) is
\begin{eqnarray}
\label{order1overN2_2}
	& & \lfac \lpfac \int
	d\Omega_{K}d\Omega_{K'} {\cal P}(\mu_{K}) {\cal P}(\mu_{K'})
	{1 \over n^{4}V_{w}^{4}}\\ \nonumber & \times& 
	\sum_{i,j,k,l} {w(\vec{s}_{i})w(\vec{s}_{j})w(\vec{s}_{k})
	w(\vec{s}_{l}) \over
	\phi(\vec{s}_{i})\phi(\vec{s}_{j})\phi(\vec{s}_{k})\phi(\vec{s}_{l})}
	e^{i\vec{K}\cdot(\vec{s}_{i}-\vec{s}_{j})}
	e^{i\vec{K}'\cdot(\vec{s}_{k}-\vec{s}_{l})} \\ \nonumber &\times &
	 \left[
	 \delta_{ij}\delta_{jk}(\langle N_i N_l \rangle -
	\langle N_i \rangle \langle N_l \rangle )
	 + 
	 \delta_{ij}\delta_{jl}(\langle N_i N_k \rangle -
	\langle N_i \rangle \langle N_k \rangle) \right. \\ \nonumber 
	 &+& \left.
	 \delta_{ik}\delta_{kl}(\langle N_i N_j \rangle -
	\langle N_i \rangle \langle N_j \rangle) 
	 + 
	 \delta_{jk}\delta_{kl}(\langle N_i N_j \rangle -
	\langle N_i \rangle \langle N_j \rangle) 
	 \right].
\end{eqnarray}
Defining $\tilde{W}_\ell^{(3,1)}$ according to eq.\ (\ref{Wnpq}),
it is straightforward to show that
\begin{equation}
\label{I3_1final}
	I_{3,1} = \delta_{\ell 0} {1\over N^2} { 4\pi \over 2\ell+1 } 
	\int {k^2 dk  \over (2\pi)^3} \tilde{W}^{(3,1)}_{\ell'}(K',k) 
	P_{\ell'}(k).
\end{equation}
The term $I_{3,2} =  I_{3,1}$ while the terms $I_{3,3}$ and $I_{3,4}$
are equivalent to $I_{3,1}$ under the transformation $\ell \leftrightarrow
\ell'$ and $K \leftrightarrow K'$.  The final remaining term in eq.\
(\ref{SIG1}) can be evaluated trivially, producing
\begin{equation}
\label{I4final}
	I_4 = {1\over N^3} \delta_{\ell 0}\delta_{\ell' 0} \int {d^3s \over 
	V_w}
	w^4(\vec{s}).
\end{equation}	
Putting all of these 
(eq.\ (\ref{I1_1final},\ref{I1_2final},\ref{I2_1final},\ref{I2_2Afinal}
,\ref{I2_2Bfinal},\ref{I3_1final},\ref{I4final})
yields eq.\ (\ref{shotnoise}) .

\newpage
\section*{Figure Captions}

\figcaption{\label{compare} Comparing the redshift evolution of 
$\bar{p}_2/\bar{p}_0$ for different cosmological models.  In the upper
panel, the starred data points show our fiducial flat model
($\lambda_0 = 0.7, b_0 = 1$) with error estimates compared with the
square data points which represent a $\lambda_0 = 0.48,~b_0 = 1.2 $ model.
Data for windows of $20 h^{-1}$ Mpc (lower curves) and $40 h^{-1}$ Mpc
(upper curves) are shown.  The comparison model has the smallest value of
$\lambda_0$ which is still acceptable at the 95\% confidence limit.
In the lower panel, the starred data represents our fiducial open
model ($\Omega_0 = 0.3$) compared with an $\Omega_0 =0.85,~b_0 = 1.6$
model, which is again just acceptable at the 95\% confidence limit.  }

\figcaption{\label{O.3L.7flat} The 68\%, 95\%, and 99\% 
confidence limits for $\Omega_0$ and $b_0$ with $\sigma$
unconstrained, when comparing flat, cosmological constant models with
a fiducial $\Omega_0 = 0.3$, $\lambda_0=0.7$, $\sigma = 300$ km/s, and
$b_0=1$ model.  The dashed curve plots $\Omega_0^{0.6}/b_0 = .5$, the
naive degeneracy expected}

\figcaption{\label{O.3L0curve}The 68\%, 95\%, and 99\% 
confidence limits for $\Omega_0$ and $b_0$ with $\sigma$
unconstrained, when comparing open  models with
a fiducial $\Omega_0 = 0.3$, $\lambda_0=0$, $\sigma = 300$ km/s, and
$b_0=1$ model.  The dashed curve plots $\Omega_0^{0.6}/b_0 = .5$, the
naive degeneracy expected}

\figcaption{\label{O.3L0flat} The 68\%, 95\%, and 99\% 
confidence limits for $\Omega_0$ and $b_0$ with $\sigma$
unconstrained, when comparing flat, cosmological constant  models with
a fiducial $\Omega_0 = 0.3$, $\lambda_0=0$, $\sigma = 300$ km/s, and
$b_0=1$ model.  The dashed curve plots $\Omega_0^{0.6}/b_0 = .5$, the
naive degeneracy expected}

\end{document}